\documentclass[preprint,12pt]{elsarticle}

\usepackage{amssymb}
\usepackage{amsmath,bm}
\usepackage[dvipsnames]{xcolor}
\journal{Journal of non-Newtonian Fluid Mechanics}

\begin{document}

\begin{frontmatter}



\title{Creeping thermocapillary motion of a Newtonian droplet suspended in a viscoelastic fluid}


\author[inst1]{Paolo Capobianchi}
%\cortext[cor2]{paolo.capobianchi@strath.ac.uk}

\affiliation[inst1]{organization={James Weir Fluid Laboratory, Department of Mechanical and Aerospace Engineering, University of Strathclyde},%Department and Organization
            addressline={75 Montrose Street}, 
            city={Glasgow},
            postcode={G1 1XJ}, 
            country={UK}}

%\affiliation[inst2]{organization={Schlumberger Cambridge Research Centre},%Department and Organization
 %           addressline={Madingley Rd}, 
  %          city={Cambridge},
   %         postcode={CB3 0EL}, 
    %        country={UK}}
            
\author[inst3]{Mahdi Davoodi}
\author[inst3]{Robert J. Poole}
\author[inst1]{Marcello Lappa}
\author[inst4]{Alexander Morozov\corref{cor1}}
\cortext[cor1]{alexander.morozov@ed.ac.uk}
\author[inst1]{ M\'{o}nica S. N. Oliveira\corref{cor2}}
\cortext[cor2]{monica.oliveira@strath.ac.uk}

\affiliation[inst3]{organization={School of Engineering, University of Liverpool},%Department and Organization
            city={Liverpool},
            country={UK}}

\affiliation[inst4]{organization={SUPA, School of Physics and Astronomy, The University of Edinburgh, James Clerk Maxwell Building},%Department and Organization
            addressline={Peter Guthrie Tait Road}, 
            city={Edinburgh},
            postcode={EH9 3FD}, 
            country={UK}}

\begin{abstract}
In this work we consider theoretically the problem of a Newtonian droplet moving in an otherwise quiescent infinite viscoelastic fluid under the influence of an externally applied temperature gradient. The outer fluid is modelled by the Oldroyd-B equation, and the problem is solved for small Weissenberg and Capillary numbers in terms of a double perturbation expansion. 

We assume microgravity conditions and neglect the convective transport of energy and momentum. We derive expressions for the droplet migration speed and its shape in terms of the properties of both fluids. In the absence of shape deformation, the droplet speed decreases monotonically for sufficiently viscous inner fluids, while for fluids with a smaller inner-to-outer viscosity ratio, the droplet speed first increases and then decreases as a function of the Weissenberg number. For small but finite values of the Capillary number, the droplet speed behaves monotonically as a function of the applied temperature gradient for a fixed ratio of the Capillary and Weissenberg numbers. We demonstrate that this behaviour is related to the polymeric stresses deforming the droplet in the direction of its migration, while the associated changes in its speed are Newtonian in nature, being related to a change in the droplet's hydrodynamic resistance and its internal temperature distribution. When compared to the results of numerical simulations, our theory exhibits a good predictive power for sufficiently small values of the Capillary and Weissenberg numbers.
\end{abstract}




\end{frontmatter}


\section{Introduction}
\label{sec:introduction}

Motion of bubbles and droplets through fluids with various rheological properties is one of the fundamental problems of fluid mechanics. Its cornerstone is the work of Hadamard-Rybczy\'{n}sky \citep{HADAMARD1911,rybczynski1911fortschreitende}, who calculated analytically the speed of a buoyancy-driven  droplet of a Newtonian fluid moving inside another Newtonian fluid. The Hadamard-Rybczy\'{n}sky theory, which assumed a spherical droplet shape and neglected inertia, has since been extended to deformable droplets \citep{taylor_acrivos_1964,pozrikidis_1990,kojima1984,koh1989}, to include the effects of inertia (e.g., see \cite{clift1978bubbles} and references therein), and to study thermocapillary migration of bubbles and droplets in an external temperature gradient \citep{young_goldstein_block_1959,SUBRAMANIAN1983145,BALASUBRAMANIAM1987531,HAJHARIRI1990277,Balasubramaniam2000,capobianchi2017}.

Our understanding of the viscoelastic analogue of the same problem is significantly less developed, and mostly limited to 
experimental (see, e.g., \cite{Philippoff1937,Warshay1959,Mhatre1959,Astarita1965,Calderbank1970,hassager1979,Liu1995}) and numerical studies (see, e.g., \cite{noh1993,WAGNER2000227,pillapakkam_singh_blackmore_aubry_2007,Mukherjee2011,CAPOBIANCHI20198}). While providing detailed information on the structure of the flow, shape of the droplet, and its migration speed, such studies seldom lead to a physical insight into the interplay of viscoelasticity, surface tension, and the externally applied driving force. Such insight can more readily be obtained from an analytical study of the same problem. When one of the fluids is viscoelastic, however, analytical analysis is often made impossible by the strongly non-linear nature of the corresponding equations of motion, and instead one has to resort to studying weakly viscoelastic fluids. Important analytical results have thus been obtained by employing various perturbation expansion techniques for the case of bubbles and droplets undergoing buoyancy driven motion in weakly viscoelastic fluids (see, e.g., \cite{Wagner1971,ajayi1975,Zana1978,TIEFENBRUCK1980257,QUINTANA1987253,CHILCOTT1988381,sostarecz_belmonte_2003}). Wagner et al. \cite{Wagner1971} calculated the drag force exerted on the droplet moving through a Rivlin-Ericksen fluid of grade three and the droplet's shape. Tiefenbruck and Leal \cite{TIEFENBRUCK1980257} focused on the motion of a spherical gas bubble through a viscoelastic fluid deriving an extended version of the Hadamard-Rybczy\'{n}sky result for the terminal velocity. 
Quintana et al. \cite{QUINTANA1987253} found that the translational velocity of a droplet can be enhanced or hindered relative to the Hadamard-Rybczy\'{n}sky value according to the degree of shear thinning, and elongational and memory effects in the viscoelastic fluid. More specifically, it was found that for large droplet viscosities, shear thinning and fluid memory cause an increase in the velocity, whereas for very mobile droplet surfaces (i.e. for gas bubbles) the motion can be accelerated or slowed down with respect to the Hadamard-Rybczy\'{n}sky value depending on the relative influence of the memory and the elongational properties of the viscoelastic phase. Chilcott and Rallison \cite{CHILCOTT1988381} extended these studies to account for the finite extensibility of the polymer molecules in the viscoelastic phase, while \cite{sostarecz_belmonte_2003} investigated experimentally and analytically the steady shape of a dilute polymer solution droplet falling through a quiescent viscous Newtonian fluid. Remarkably, \cite{sostarecz_belmonte_2003} demonstrated that the dimpled shape displayed by the droplet under some conditions could be reproduced analytically by considering the axisymmetric Stokes flow past a non-Newtonian drop, modelled as a Simple Fluid of Order Three. 

To the best of our knowledge, at the time of preparation of this work \cite{capobianchi2020creeping} the only analytical study about Marangoni migration of droplets in the presence of viscoelastic effects was that of \cite{jimenez_crespo}, who considered the steady thermocapillary motion of an inviscid spherical bubble under gravity. Neglecting the momentum and convective heat transfer, \cite{jimenez_crespo} modelled the outer fluid by the Oldroyd-B constitutive equation, restricting their analysis to weak viscoelasticity only. Their results provide the force exerted by the fluid on the bubble and a weakly viscoelastic correction to its terminal velocity originally derived for pure Marangoni and mixed buoyancy-thermocapillary flows by \cite{young_goldstein_block_1959}. We should note however that since then, Vyas and Ghosh \cite{vyasetal} attempted to analyse the effect of an imposed temperature gradient on the motion of a viscoelastic drop suspended in another viscoelastic medium being driven by an external pressure gradient using asymptotic analysis. The authors consider a Phan-Thien Tanner constitutive model for both fluids and predict that the viscoelastic properties of the droplet strongly influence the drop deformation and migration velocity. They also predict a maximum migration velocity for an intermediate viscosity of the interior phase for Marangoni numbers larger than zero (provided the droplet exhibits has stronger viscoelasticity than the suspending medium), while for negative Marangoni numbers, the drop's motion can be completely arrested.  Drop shape changes from prolate to oblate are reported when viscoelastic stresses become significant. 

In this work, we extend the previous study of \cite{jimenez_crespo} and consider the migration of deformable Newtonian droplets in a weakly viscoelastic fluid in the presence of thermal Marangoni effects. By neglecting momentum and convective heat transfer, and  assuming microgravity conditions, we develop a perturbation theory to study analytically the combined effects of non-uniform thermal distribution, leading to interfacial tension gradients, and viscoelasticity. We show that viscoelasticity has significant implications for the velocity of the droplet, which can either increase or decrease as a function of the applied temperature gradient depending on the ratios of the viscosities and thermal conductivities of the two fluids. We also demonstrate that viscoelastic stresses universally stretch the droplet along the direction of its migration, leading to fore-aft asymmetric shapes.

The remainder of the paper is organised as follows. In \S \ref{sec:problem_statement} we state the problem we study, while in \S \ref{sect:mathematical_formulation} we provide the equations of motion and the boundary conditions. In \S \ref{sect:solution_procedure} we outline the procedure used to solve the equations of motion perturbatively and derive the migration speed and the shape of the droplet. In \S \ref{sect:discussion} we discuss the physical implications of our results and compare them to the numerical simulations of \cite{CAPOBIANCHI20198}, while in \S\ref{sect:conclusions} we present our conclusions.

\section{Statement of the problem}
\label{sec:problem_statement}

We consider a Newtonian droplet moving in an infinite viscoelastic fluid. The two fluids are assumed to be perfectly immiscible. Although we refer to the inner phase as the \emph{droplet}, the analysis presented below also applies to the case of a gas bubble.  The droplet is assumed to be deformable, and has the volume of an equivalent sphere of radius $R$. The outer viscoelastic phase is characterised by a single Maxwell relaxation time $\lambda$, and a constant viscosity, $\eta_0 = \eta_s + \eta_p$, where $\eta_s$ and $\eta_p$ and the Newtonian (solvent) and polymeric contributions, respectively, which has been extensively used to model polymeric viscoelastic solutions of constant viscosity \cite{SHAQFEH2021104672} commonly known as Boger fluids. The droplet phase has the viscosity $\tilde\eta$. In the following, tildes are used to denote quantities referring to the inner phase.

\begin{figure}
\centering
\includegraphics[width=0.4\textwidth]{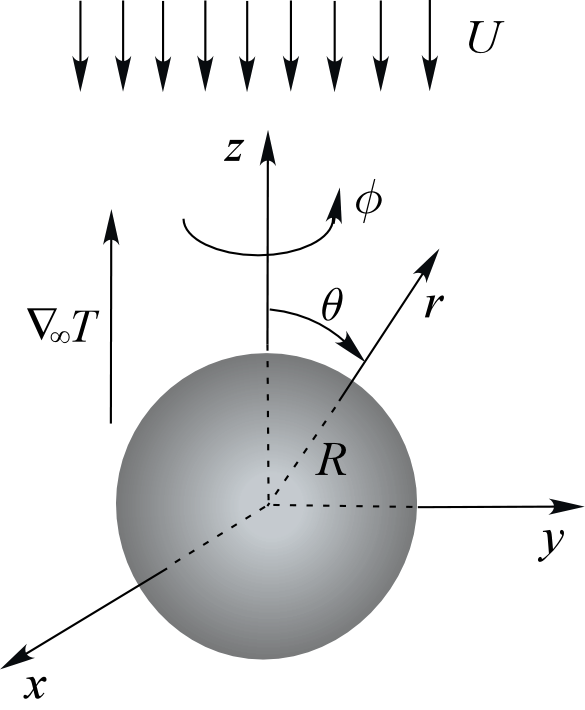}
\caption{ Schematic of a droplet in the presence of a temperature gradient, ${\nabla}_{\infty}T$. The centre of mass of the droplet is chosen as the origin of a spherical coordinate system. The fluid at infinity moves with the velocity $-U$ in the direction opposite to the applied temperature gradient. }
\label{fig:drop_scheme}
\end{figure}

The motion is generated by a constant temperature gradient, ${{\nabla} _\infty }T$, which is maintained by external means (see Fig. \ref{fig:drop_scheme}). We assume that the interfacial tension $\sigma$ between the inner and the outer fluids is a decreasing function of the temperature with a constant gradient ${{\sigma }_{T}}={\partial\sigma}/{\partial T} < 0$ \citep{Dee1999}. 
Furthermore, gravity is neglected throughout this work. Under these assumptions, the droplet migrates along the direction of the imposed thermal gradient by virtue of thermocapillary effects \citep{young_goldstein_block_1959}. Following common practice (see, for instance, \cite{subr_balasubr:2001}), we assume that the remaining fluid properties are insensitive to temperature variations, and that the droplet attains a steady-state velocity, $U$. We stress that this assumption can be violated under certain conditions, especially in the case of the viscosity and relaxation time of non-Newtonian fluids in the presence of large temperature gradients (see, e.g., \cite{Rothstein2001}), and, thus, care is needed when comparing our results to experiments.

To proceed, we employ a spherical coordinate system, $(r,\theta,\phi)$, with its origin being fixed at the centre of mass of the translating droplet, see Fig.\ref{fig:drop_scheme}. In this coordinate system, the droplet is stationary, while the fluid at infinity moves with the velocity $-U$ in the direction opposite to the applied temperature gradient.

The problem is rendered dimensionless by adopting $R$ as a reference length, $U_T=-\sigma_T \nabla_\infty T R/\eta_0$ as the velocity scale \citep{subr_balasubr:2001}, and the convective timescale $t_c=R/U_T$ as the characteristic time. Stresses and the pressure in the outer phase are normalised with the characteristic viscous stress ${{\eta }_{0}}{{U}_{T}}/R$, while ${\tilde{\eta }}{{U}_{T}}/R$ is used in the Newtonian phase. A dimensionless temperature is defined as $\vartheta({\bm r}) = (T({\bm r})-T_0)/ R \nabla_\infty T$, where $T({\bm r})$ is a local temperature and $T_0$ is a reference temperature, chosen to be the temperature of the unperturbed linear profile at the current position of the droplet's centre of mass; in these units the temperature profile in the absence of the droplet is given by $\vartheta({\bm r}) = z = r \cos{\theta}$, see Fig.\ref{fig:drop_scheme}. The value of the surface tension at the reference temperature, $\sigma(T_0)$, is used as a scale for the surface tension, leading to the following dimensionless profile: $\sigma({\bm r}_i) = 1 - U_T \eta_0 \vartheta({\bm r}_i)/\sigma(T_0)$, where ${\bm r}_i$ is a position at the interface; since the temperature is continuous across the interface, there is no need to distinguish between $\vartheta({\bm r}_i)$ and $\tilde\vartheta({\bm r}_i)$ in this equation. All quantities presented below are dimensionless, unless explicitly stated otherwise.

The problem is characterised by the following dimensionless numbers. The importance of inertia over the viscous stresses is determined by the Reynolds number, defined as $Re=\rho R U_T/\eta_0$, where $\rho$ is the density of the outer phase. 
The strength of heat advection set in motion by the surface tension gradients as compared to thermal diffusion is determined by the Marangoni number, $Ma=R{{U}_{T}}/K$, where $K$ is the thermal diffusivity of the outer phase.  The ratio of the typical magnitudes of viscous stresses and surface tension is given by the Capillary number, $Ca={{\eta }_{0}}{{U}_{T}}/\sigma(T_0)$, that determines whether the droplet can be significantly deformed by the ensuing motion. Finally, the magnitude of elastic stresses is controlled by the Weissenberg number, $Wi=\lambda U_T/R$. 

In what follows, we assume that both Reynolds and Marangoni numbers are sufficiently small, and we neglect convective transport of energy and momentum in the equations of motion. 

\section{Equations of motion and boundary conditions}
\label{sect:mathematical_formulation}

Employing the dimensionless units introduced in Section \ref{sec:problem_statement}, the equations of motion for the viscoelastic outer phase read
\begin{align}
\label{eqn:10}
& \bm{\nabla}  \cdot \bm{u} = 0,\\
\label{eqn:11}
& \bm{\nabla} p = \bm{\nabla}  \cdot \bm{\tau},\\
\label{eqn:12}
& {\bm{\nabla} ^2}\vartheta = 0,
\end{align}
while the governing equations for the Newtonian droplet phase are given by
\begin{align}
\label{eqn:13}
& \bm{\nabla}  \cdot \tilde{\bm{u}}  = 0,\\
\label{eqn:14}
& \bm{\nabla} \tilde p = \bm{\nabla}  \cdot \widetilde{\textbf{D}},\\
\label{eqn:15}
&{\bm{\nabla} ^2}\widetilde \vartheta = 0.
\end{align}
Here, $\bm{u}$ and $\tilde{\bm{u}}$, $p$ and $\tilde p$, and $\vartheta$ and $\widetilde \vartheta$, are the dimensionless velocities, pressures, and temperatures of the outer and inner fluids, respectively;  $\widetilde{\textbf{D}} = {\bm\nabla} \tilde{\bm u} +\left( {\bm\nabla} \tilde{\bm u} \right)^{\rm{T}}$, where $\rm{T}$ denotes transpose of a matrix. The total dimensionless deviatoric stress in the outer fluid, $\bm{\tau}$, comprises both viscous and viscoelastic contributions, and is assumed to satisfy the viscoelastic Oldroyd-B model \citep{oldroyd1950,bird1987dynamics}
\begin{equation}
\bm{\tau}  + Wi\, \hat d\bm{\tau}  = \textbf{D} + \beta Wi \hat d\textbf{D},
\label{eq:oldroyd}
\end{equation}
where $\beta = \eta_s/\eta_0$ is the ratio between the solvent and the total viscosities of the outer phase, $\textbf{D} = \bm{\nabla} \bm{u} + {\left( {\bm{\nabla} \bm{u}} \right)^{\rm{T}}}$, and 
\begin{equation}
\hat d\left( \cdot \right) = {{d( \cdot )} \mathord{\left/
 {\vphantom {{d( \cdot )} {dt}}} \right.
 \kern-\nulldelimiterspace} {dt}} 
 + {\bm{u}} \cdot \bm{\nabla} \left(  \cdot  \right) - {\bm{\nabla} {\bm{u}}^{\rm{T}} \cdot \left(  \cdot  \right) - \left(  \cdot  \right) \cdot \bm{\nabla} {\bm{u}}} 
 \label{eq:upper-convected}
\end{equation}
defines the upper-convected derivative operator.

The velocity, pressure, and temperature fields in the outer and inner fluids are related through a set of boundary conditions specified at the droplet's interface. In the following, we assume that all fields are axisymmetric, i.e. no quantity depends on the azimuthal angle $\phi$. In this case, the position of the interface can be defined as \citep{taylor_acrivos_1964}
\begin{align}
r = 1 + \zeta \left( \theta  \right),
\label{eq:interfacedefinition}
\end{align}
where $\zeta \left( \theta  \right)$ is an unknown function to be determined as a part of the solution. Eq.\eqref{eq:interfacedefinition} allows us to introduce the local normal
\begin{align}
{\bm n} = \frac{\left(1+\zeta(\theta),-\zeta'(\theta),0\right)}{\sqrt{\left(1+\zeta(\theta)\right)^2+\zeta'(\theta)^2}}
\end{align}

and tangential
\begin{align}
{\bm t} = \frac{\left(\zeta'(\theta),1+\zeta(\theta),0\right)}{\sqrt{\left(1+\zeta(\theta)\right)^2+\zeta'(\theta)^2}}
\end{align}

vectors to the interface, where primes denote derivatives w.r.t. $\theta$. For a spherical droplet, $\zeta(\theta)=0$, and the equations above trivially reduce to ${\bm n}=\hat{\bm e}_r$ and ${\bm t}=\hat{\bm e}_\theta$, where $\hat{\bm e}_r$ and $\hat{\bm e}_\theta$ are the corresponding unit vectors of our spherical coordinate system.

The boundary conditions employed in this work are given by the following set of equations
\begin{align}
{\bm u}\cdot {\bm n} = &\,\tilde{\bm u}\cdot {\bm n} = 0, \label{eq:velkinematic}\\
{\bm u}\cdot {\bm t} &= \tilde{\bm u}\cdot {\bm t}, \label{eq:velcont}\\
\vartheta &= \tilde \vartheta,  \label{eq:tempcont}\\
{\bm n}\cdot {\bm \nabla} \vartheta &= \gamma {\bm n}\cdot {\bm \nabla} \tilde \vartheta,  \label{eq:tempflux}\\
\alpha\,\tilde p - p + \bm{n} \cdot \left( {\bm{\tau}  - \alpha \, \widetilde{\textbf{D}} } \right)& \cdot \bm{n}  = \frac{1}{{Ca}}\left( {1 - Ca\,\vartheta} \right)\bm{\nabla}  \cdot \bm{n}, \label{eq:stressnormal}\\
\bm{t} \cdot \left( {\bm{\tau}  - \alpha \, \widetilde{\textbf{D}} } \right)& \cdot \bm{n} = \bm{t} \cdot \bm{\nabla} \vartheta,
\label{eq:stresstangential}
\end{align}
where all quantities are evaluated at the position of the interface, $r=1+\zeta(\theta)$. Here, $\alpha=\tilde{\eta}/\eta_{0}$ is the viscosity contrast between the inner and outer fluids, and $\gamma$ denotes the ratio of the thermal conductivity of the droplet phase to that of the suspending fluid. It should be emphasised that the term $\vartheta \bm{\nabla}  \cdot \bm{n}$ appearing in Eq. (16) represents a non-linear coupling between the shape of the droplet and the temperature profile, making it impossible to solve the equations of motion exactly at each order in $Wi$.
The kinematic conditions, Eqs.\eqref{eq:velkinematic}, imply that the interface is stationary, while Eqs.\eqref{eq:velcont}, \eqref{eq:tempcont} and \eqref{eq:tempflux} ensure that the tangential components of the velocity, temperature, and the temperature flux are continuous across the interface. The final conditions, Eqs.\eqref{eq:stressnormal} and \eqref{eq:stresstangential}, ensure the balance of the normal and tangential components of the local force per unit area acting on the interface; in writing these equations, we have taken into account that the interfacial tension is a function of the local temperature  (see, e.g., \cite{subr_balasubr:2001}).

Far away from the droplet, the velocity and temperature fields must assume their unperturbed values
\begin{align}
{\bm u} \to -U \left( \hat{\bm e}_r\cos{\theta} -  \hat{\bm e}_\theta \sin{\theta}\right), \quad \mathrm{and} \quad \vartheta \to r\cos{\theta}, \quad \mathrm{as} \quad r\to\infty,
\label{eq:VelAtInfinity}
\end{align}
where $U$ is the yet to be determined droplet speed, cf. Fig. \ref{fig:drop_scheme}. All fields must also be regular inside the droplet. Furthermore, the total force acting on a neutrally buoyant droplet should be zero, with only the $z$-component of the force providing a non-trivial condition. Formulated for the inner phase, this requirement yields
\begin{align}
& \int_0^{\pi} d\theta \sin{\theta}\big(1+\zeta(\theta)\big)\sqrt{\big(1+\zeta(\theta)\big)^2 + \zeta'^2(\theta)} \nonumber \\
& \qquad\qquad\qquad\qquad\qquad \times \left( \hat{\bm e}_r\cos{\theta} -  \hat{\bm e}_\theta \sin{\theta}\right) \cdot \left[ -\tilde p \, \textbf{I} + \widetilde{\textbf{D}} \right] \cdot {\bm n} = 0,
\label{eq:Fz}
\end{align}
where $\textbf{I}$ is the identity tensor. 

Finally, the shape deformation given by Eq.\eqref{eq:interfacedefinition} should satisfy two constraints. The first is given by the droplet volume conservation
\begin{equation}
\label{eq:interface_constraint_1}
\frac{1}{2}\int_{0}^\pi d\theta \sin\theta \big[ 1 + \zeta(\theta) \big]^3 = 1,
\end{equation}
while the second stipulates that the position of the droplet's centre of mass is not changed by the deformation, yielding
\begin{equation}
\label{eq:interface_constraint_2}
\int_{0}^\pi d\theta\, \sin2\theta \big[ 1 + \zeta(\theta) \big]^4 = 0.
\end{equation}

\section{Solution procedure}
\label{sect:solution_procedure}

\subsection{Newtonian solution}
\label{sect:newtonian_solution}

Before proceeding to discuss our treatment of the viscoelastic problem defined in Eqs.\eqref{eqn:10} - \eqref{eq:Fz}, here we briefly review the solution of its Newtonian analogue \citep{young_goldstein_block_1959}, obtained by setting $Wi=0$ in Eq.\eqref{eq:oldroyd}. In this case, the stress tensor in the outer fluid reduces to the viscous contribution, $\bm{\tau} ^{(0,0)} = \textbf{D}^{(0,0)}$, and the velocity, pressure, and temperature in both fluids satisfy the same system of linear homogeneous equations; here, we use the superscript $(0,0)$ to denote the Newtonian solution.  

For an incompressible axisymmetric flow, it is convenient to re-formulate the problem in terms of the Stokes streamfunctions $\psi \left( r, \theta \right)$  and  $\widetilde\psi \left( r, \theta \right)$ \citep{happel2012low}, defined through
\begin{align}
\label{eqn:47}
{u_{r}} =  - \frac{1}{{{r^2}\sin \theta }}\frac{{\partial {\psi}}}{{\partial \theta }}, \quad {u _{\theta} } = \frac{1}{{r\sin \theta }}\frac{{\partial {\psi}}}{{\partial r}},
\end{align}
and similar expressions for the inner fluid; Eqs.\eqref{eqn:47} ensure that the incompressibility condition is satisfied by definition \citep{happel2012low}. In this formulation, Stokes equations Eqs.\eqref{eqn:11} and \eqref{eqn:14} reduce to two identical problems in the Newtonian limit, ${E^4}\psi^{(0,0)}  = 0$ and ${E^4}\widetilde \psi^{(0,0)}  = 0$, where
\begin{align}
{E^4} = \left( {\frac{{{\partial ^2}}}{{\partial {r^2}}} + \frac{{\sin \theta }}{{{r^2}}}\frac{\partial }{{\partial \theta }}\left( {\frac{1}{{\sin \theta }}\frac{\partial }{{\partial \theta }}} \right)} \right)^2
\end{align}
is the bi-harmonic operator in spherical coordinates. Their general solution, regular for any $\theta$, is given by \citep{happel2012low}
\begin{align}
\label{eqn:streamfunction_solution}
& \psi \left( r,\theta \right) \nonumber \\
& \,\, = \sum\limits_{n = 2}^\infty  { \left( {{C^{(1)}_n}{r^{n}} + {C^{(2)}_n}{r^{- n+1}} + {C^{(3)}_n}{r^{n + 2}} + {C^{(4)}_n}{r^{-n + 3}}} \right)} \frac{P_{n-2}(\cos\theta) - P_{n}(\cos\theta)}{2n-1},
\end{align}
and similar for the inner fluid. Here, $P_n$ is the Legendre polynomial of degree $n$, and
$C^{(1)}_n,\dots,C^{(4)}_n$  are unknown constants. In a similar fashion, the relevant solution of the Laplace equation for the dimensionless temperature is given by \citep{subr_balasubr:2001}
\begin{equation}
\label{eqn:temperature_solution}
\vartheta\left( {r,\theta} \right) = \sum\limits_{n = 0}^\infty  \left( {{C^{(5)}_n}{r^n} + {C^{(6)}_n}{r^{ - n - 1}}} \right) {P_n}\left( \cos\theta \right) ,
\end{equation}
and a similar series for $\tilde\vartheta$; again, $C^{(5)}_n$ and $C^{(6)}_n$ are unknown constants. 

The constants in the general solutions presented above are set by the conditions Eqs.\eqref{eq:velkinematic}-\eqref{eq:Fz}. For a general shape of the droplet, however, these lead to strongly non-linear equations for the unknown constants that have no closed-form analytical solution. In the case where both fluids are Newtonian, however, a significant simplification occurs. Indeed, setting $\zeta(\theta)=0$, we observe that 
\begin{align}
& \vartheta^{(0,0)}\left( {r,\theta} \right) = \left( {r + \frac{{1 - \gamma }}{{2 + \gamma }}\frac{1}{{{r^2}}}} \right)\cos\theta,
\qquad 
\widetilde \vartheta^{(0,0)}\left( {r,\theta} \right) = \frac{{3r}}{{2 + \gamma }}\cos\theta,
\label{eq:TNewt} 
\\
& {\psi^{(0,0)}}\left( {r,\theta} \right)  = A\left( {{r^2} - \frac{1}{r}} \right)\sin^2\theta, 
\qquad\quad 
{\widetilde \psi^{(0,0)}}\left( {r,\theta} \right)  = \frac{3}{2}A\left( {{r^4} - {r^2}} \right)\sin^2\theta, 
\label{eq:PsiNewt} 
\\
& p^{(0,0)}\left( {r,\theta} \right) = 0, \qquad\qquad\qquad\qquad\qquad\quad \tilde p^{(0,0)}\left( {r,\theta} \right) = \frac{2}{\alpha\,Ca} - 30 A\,r\cos \theta
\end{align}
satisfy all equations and boundary conditions Eqs.\eqref{eqn:10} - \eqref{eq:Fz}. Here, 
\begin{align}
A = \frac{1}{(2+\gamma)(2+3\alpha)},
\label{eq:Adefinition}
\end{align}
and we have set the irrelevant constant pressure at infinity to zero. We can, therefore, conclude that in the absence of convective transport of momentum and energy, mechanical stresses generated in the fluid by the ensuing motion do not deform the droplet from the spherical shape. Finally, we observe that Eqs.\eqref{eq:PsiNewt} and \eqref{eqn:47} imply that 
far away from the droplet ${\bm u} \to -2A \left( \hat{\bm e}_r\cos{\theta} - \hat{\bm e}_\theta \sin{\theta}\right)$. Comparing this result with Eq.\eqref{eq:VelAtInfinity}, we obtain the Newtonian droplet speed $U^{(0,0)} = 2A$.

\subsection{Asymptotic expansion}
\label{sect:asymptotic_expansion}
To simultaneously account for viscoelasticity of the outer fluid and a finite droplet deformability, we seek the solution to the general problem defined in Eqs.\eqref{eqn:10} - \eqref{eq:Fz} in the form of a double expansion
\begin{align}
X = \sum_{\substack{n=0\\m=0}}^\infty X^{(n,m)}Wi^n Ca^m,
\label{eqn:general_expansion}
\end{align}
where $X$ stands for the velocity, pressure, stress, and temperature fields in both fluids, as well as for the shape function $\zeta(\theta)$ and the droplet speed $U$; this form implies that we assume $Wi < 1$ and $Ca < 1$. As discussed in Section \ref{sect:newtonian_solution}, no droplet deformation occurs in the Newtonian limit $Wi=0$, hence $X^{(0,m)}=0$ for $m>0$, while $X^{(0,0)}$ refer to the Newtonian values for the respective quantities presented above.

For a given order in the expansion, we utilise the following solution procedure. Expanding Eq.\eqref{eq:oldroyd} to the corresponding order in $Wi$ and $Ca$, we observe that it becomes an algebraic equation for the unknown stress components. To illustrate this, we consider $O(Wi)$, which gives
\begin{align}
& {\bm{\tau}^{(1,0)}}  = {\textbf{D}^{(1,0)}} \nonumber \\
& \qquad\qquad - \left( {1 - \beta } \right)
\Big[ 
{\bm{u}^{(0,0)}} \cdot \bm{\nabla} {\textbf{D}^{(0,0)}} - \left({\bm{\nabla}} {\bm{u}^{(0,0)}}\right)^{\rm{T}} \cdot {\textbf{D}^{(0,0)}} - {\textbf{D}^{(0,0)}} \cdot \bm{\nabla} {\bm{u}^{(0,0)}}
\Big].
\label{eq:stress11example}
\end{align}
At higher orders in $Wi$ and $Ca$, the stress components retain the same structure, comprising a purely viscous contribution $\textbf{D}$ at the same order, and quadratic combinations of velocities and stresses at lower orders through the upper-convected derivative, Eq.\eqref{eq:upper-convected}.

To obtain the equations for the streamfunctions $\psi^{(n,m)}$ and $\widetilde \psi^{(n,m)}$, we take the curl of Eqs.\eqref{eqn:11} and \eqref{eqn:14}, and project it onto the azimuthal direction, $\bm{\hat e}_\phi$. While in the Newtonian phase the resulting equation is still homogeneous, ${E^4}{\widetilde \psi^{(n,m)}} = 0$, the streamfunction in the viscoelastic (outer) phase is now given by an inhomogeneous equation, ${E^4}{\psi^{(n,m)}} = g(r,\theta)$, where $g(r,\theta)$ originates from the upper-convected derivative terms in the solution for the stress components (\emph{e.g.} the second term in Eq.\eqref{eq:stress11example}). The solution to the outer problem consists of an inhomogeneous part, determined by $g(r,\theta)$, and a homogeneous part, given by Eq.\eqref{eqn:streamfunction_solution}, where only terms that match the angular symmetry of $g(r,\theta)$ are retained. The inner solution is then given by the modes in Eq.\eqref{eqn:streamfunction_solution} with the same angular symmetry. Some of the unknown constants of the outer solution have to be set to zero to ensure the condition far away from the droplet, Eq.\eqref{eq:VelAtInfinity}, while the inner velocity field has to be regular at $r=0$.

The pressure in both fluids is obtained by substituting the streamfunctions and stresses back into Eqs.\eqref{eqn:11} and \eqref{eqn:14}, leading to simple first-order partial differential equations that are readily solved. We stress the importance of introducing an unknown constant into the expression for the inner pressure, which will be crucial in determining the shape of the droplet. The corresponding constant term in the outer problem is set to zero since, at any order, the outer pressure should decay to the pressure in the quiescent fluid far away from the droplet.

Next, we turn to the balance of the normal force given by Eq.\eqref{eq:stressnormal}. At every order, it relates 
$\bm{\nabla}  \cdot \bm{n}$ to the already known quantities at lower orders, thus allowing us to determine the interface deformation $\zeta^{(n,m)}$. As mentioned above, the constant inner pressure contributions, undetermined by the equations at their respective order in $Wi$ and $Ca$, are set by requiring that $\zeta(\theta)$ satisfies Eqs.\eqref{eq:interface_constraint_1} and \eqref{eq:interface_constraint_2} at every order. 

Since the Newtonian temperature fields, Eqs.\eqref{eq:TNewt}, satisfy the boundary conditions, Eqs.\eqref{eq:tempcont} and \eqref{eq:tempflux}, at the unperturbed surface $r=1$, any change to the droplet's shape drives a correction to the inner and outer temperature distributions. The angular symmetry of this correction is set by the continuity of the tangential force acting locally on the interface, Eq.\eqref{eq:stresstangential}, resulting in only a few terms from the general solution, Eq.\eqref{eqn:temperature_solution}, contributing at a given expansion order. Together with Eqs.\eqref{eqn:12}, \eqref{eqn:15}, \eqref{eq:tempcont}, \eqref{eq:tempflux}, \eqref{eq:VelAtInfinity} and the regularity condition at $r=0$, this completely determines $\vartheta^{(n,m)}\left( {r,\theta} \right)$ and $\widetilde \vartheta^{(n,m)}\left( {r,\theta} \right)$.

The remaining unknown constants are trivially fixed by the boundary conditions, Eqs.\eqref{eq:velkinematic} and \eqref{eq:velcont}, and the no-force requirement, Eq.\eqref{eq:Fz}. 

The procedure outlined above can, in principle, be applied at any expansion order, although the expressions involved quickly become very cumbersome. In this work, we employ a spherical cut-off, $n+m\le3$, \emph{i.e.} we only consider contributions proportional to $Wi$, $Ca\, Wi$, $Wi^2$, $Ca^2 Wi$, $Ca\, Wi^2$, and $Wi^3$. To this accuracy, we obtain for the speed of the droplet
\begin{align}
& \frac{U}{U^{(0,0)}} = 1
+\frac{6}{25} Ca\,Wi\,A^2 (1-\beta )\frac{22+13 \alpha}{1+\alpha} \left( \frac{33-18\alpha}{3(2+3\alpha)}+\frac{3(\gamma-1)}{2+\gamma}  \right) \nonumber \\
& \qquad\qquad\quad -\frac{54}{3575} Wi^2 A^2(1-\beta)\left(5\frac{26+193 \alpha }{2+3 \alpha }+286\frac{ 1-\beta }{1+\alpha }\right),
\label{eq:speedfinal}
\end{align}
It is worth mentioning that the terms proportional to $Wi$, $Wi^3$, $Ca^2Wi$, and $Ca Wi^2$ are all zero by symmetry and thus do not appear in Eq.\eqref{eq:speedfinal}: the magnitude of the droplet's velocity should be insensitive to the reversal of the direction of the temperature gradient.
Moreover, the droplet's shape is given by
\begin{align}
& \zeta(\theta)=\frac{3}{5} Ca\,Wi\, A^2 (1-\beta ) \frac{22+13 \alpha }{1+\alpha }P_2(\cos\theta) \nonumber \\
& +\frac{27}{1750} Ca^2 Wi\,A^3 (1-\beta ) \frac{(22+13 \alpha ) \left(28 \alpha ^2+17 \alpha -10\right) }{(1+\alpha )^2}P_3(\cos\theta) \nonumber \\
& -\frac{27}{2275} Ca\,Wi^2 A^3 (1-\beta ) \nonumber \\
& \qquad\qquad\qquad\qquad\times \frac{ (1+\alpha ) (5480+4481 \alpha )-39 (1-\beta )(142+103\alpha ) }{ (1+\alpha )^2} P_3(\cos\theta).
\label{eq:shapefinal}
\end{align}
The corresponding equations for the stress, pressure, temperature, and streamfunctions in both fluids are given in Supplementary Material.

\section{Discussion}
\label{sect:discussion}

Eqs.\eqref{eq:speedfinal} and \eqref{eq:shapefinal} constitute the main results of our work. They represent the speed and the shape of a Newtonian droplet moving through a model viscoelastic medium due to the presence of an externally applied temperature gradient. Our results are obtained in the limit of weak viscoelasticity and high surface tension, and are presented in a form of an asymptotic double series in $Wi$ and $Ca$. Below we discuss their implications for the motion of the droplet, and their limits of applicability.

First, we comment on our choice of $U_T$ as a velocity scale. As discussed in Section \ref{sec:problem_statement}, $U_T$ represents a typical velocity of the fluid set in motion by a balance of the thermocapillary interfacial  and viscous stresses. 
While incorporating the relevant physical ingredients, it does not correctly capture a typical magnitude of the fluid velocity, which in the Newtonian case is given instead by $2 A\,U_T$. According to Eq.\eqref{eq:Adefinition}, $A$ does not exceed $1/4$, while its typical values are yet smaller. For instance, for the inner and outer fluids with similar viscosities and thermal conductivities, $\alpha\sim1$ and $\gamma\sim1$, Eq.(4.8) gives $A\sim 0.07$. In other words, while $U_T$ is the correct dimensional combination, it lacks a dimensionless factor $A$, which is not of order unity for this problem. 
This implies that $Wi$  significantly overestimates the effect of viscoelasticity, while $Ca$ underestimates the effect of
the capillary forces, since both are based on $U_T$, and that the right scale for these effects is set by $Wi^{(A)} = 2 A Wi$ and $Ca^{(A)} = 2 A\,Ca$. Indeed, as Eqs.\eqref{eq:speedfinal} and \eqref{eq:shapefinal} suggest, our theory is valid for small values of $Wi^{(A)}$ and $Ca^{(A)}$, while $Wi$ and $Ca$ can potentially be large.

\begin{figure}
\centering
\includegraphics[width=0.75\textwidth]{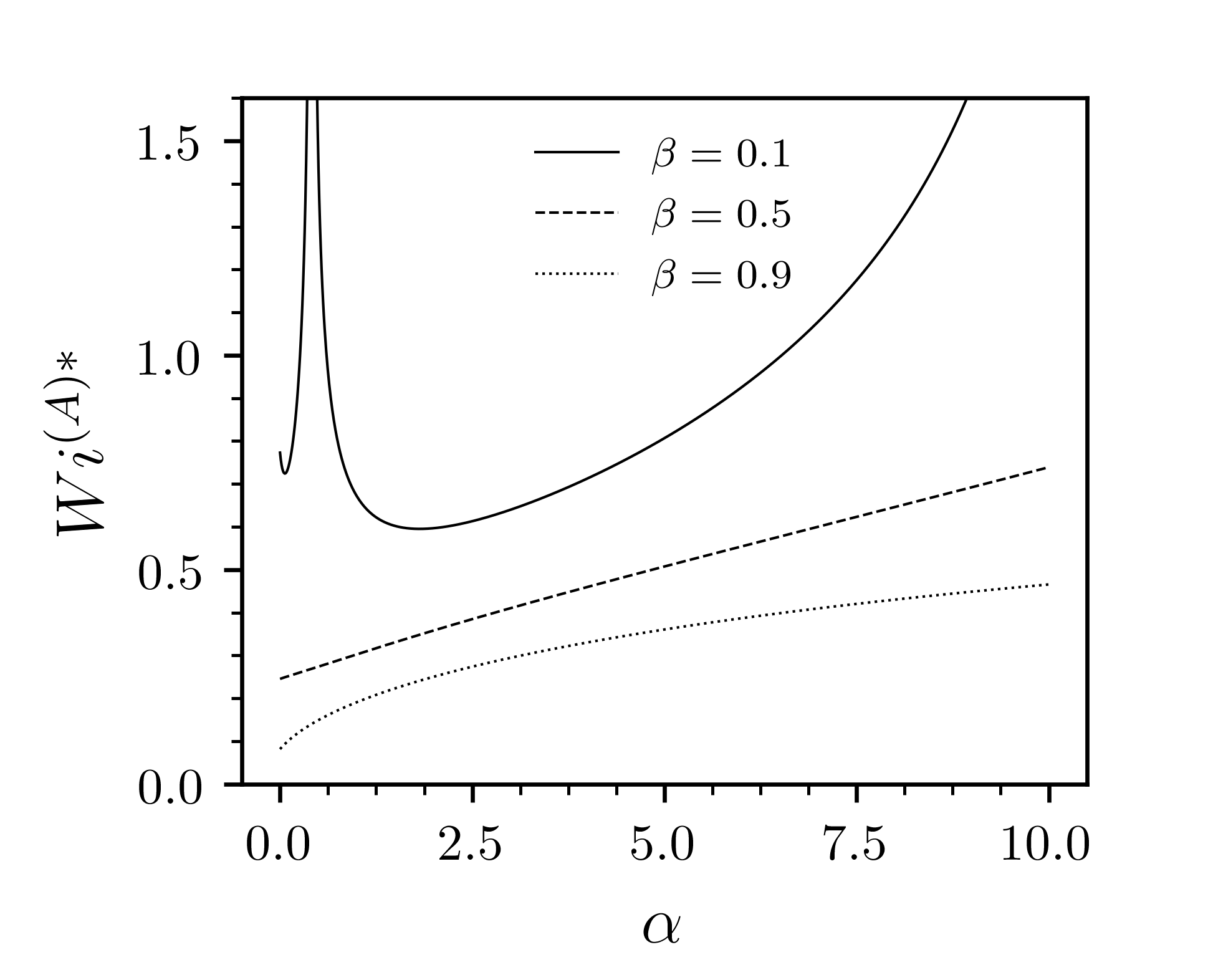}
\caption{The applicability range of our perturbation theory, $Wi^{(A)}<Wi^{(A)*}$, as a function of $\alpha$ for several values of $\beta$.}
\label{fig:maxDeT}
\end{figure}

Our results are presented in terms of power series that often have a very small, potentially zero, radius of convergence \citep{Leslie1961,Roy2006,Housiadas2011,Moore2012}. To estimate the range of their validity, we consider the case of a spherical droplet, $Ca=0$. We extend our analysis to $O(Wi^{(A) 4})$ (details not shown) and obtain the following result for the speed of the droplet
\begin{align}
& \frac{U}{U^{(0,0)}} = 1
-\frac{27}{7150}Wi^{(A) 2} (1-\beta )\Bigg(5\frac{26+193 \alpha }{2+3 \alpha }+286\frac{ 1-\beta }{1+\alpha }\Bigg) \nonumber\\
&-\frac{9}{48412} Wi^{(A) 4} (1-\beta ) \Bigg( \frac{697410+39343 \alpha }{2+3 \alpha } \nonumber \\
&\qquad -(1-\beta )\frac{17647994572+31470338748 \alpha +9825203787 \alpha ^2+3003766830 \alpha ^3}{7150(1+\alpha ) (2+3 \alpha )^2} \nonumber \\
&\qquad-(1-\beta )^2\frac{170642956+600328498 \alpha -416798595 \alpha ^2-150048675 \alpha ^3}{ 2750 (1+\alpha )^2 (2+3 \alpha )} \nonumber \\
&\qquad+1539(1-\beta )^3\frac{5412536+5092704 \alpha +2500320 \alpha ^2+376425 \alpha ^3}{13750 (1+\alpha )^3 (2+3 \alpha )}\Bigg),
\label{eq:spherical4}
\end{align}
where, again, $U^{(0,0)}=2A$. For Eqs.\eqref{eq:speedfinal} and \eqref{eq:shapefinal} to be semi-quantitatively accurate, we require the terms retained in those equations to be sufficiently larger than the next higher order term. Equating the absolute values of the second and the third terms in Eq.\eqref{eq:spherical4} gives a na\"ive estimate of $Wi^{(A)*}$ that  should not be exceeded for this condition to hold. In Fig.\ref{fig:maxDeT} we plot $Wi^{(A)*}$ as a function of the viscosity ratio $\alpha$ for several values of $\beta$. We observe that for sufficiently large values of $\beta$, $Wi^{(A)*}\sim0.4$ provides a consistent estimate for the range of validity of our theory. For $\beta=0.1$, there exist such values of $\alpha$ that the term proportional to $Wi^{(A)4}$ vanishes. Such points correspond to the divergences in Fig.\ref{fig:maxDeT} visible around $\alpha\approx0.4$ and $\alpha\approx11$ for $\beta=0.1$. Around these points the fourth-order term is very small and cannot be reliably used to estimate $Wi^{(A)*}$. Instead, one should consider the next non-vanishing term in the expansion, which goes beyond the scope of this work. Here, we choose to use a conservative condition $Wi^{(A)}  < 0.4$ across all values of parameters considered. Since our expansion, Eq.\eqref{eqn:general_expansion}, tacitly assumes that $Ca$ and $Wi$ are of the same order of smallness, a similar condition is applied to $Ca^{(A)}$.

To analyse the predictions of Eq.\eqref{eq:speedfinal} in various situations, we consider the following archetypal sets of parameters. The first set, $\alpha = 0$ and $\gamma = 0$, corresponding to a Newtonian droplet with a very low viscosity and thermal conductivity, represents a broad class of gas bubbles suspended in viscoelastic solutions. Next, we observe that at room temperature the thermal conductivities of various Newtonian and polymeric liquids are quite similar \citep{Broniarz2009}. Therefore, the other three sets of parameters we are going to study below are $\gamma=1$ and $\alpha=0.1$, $\gamma=1$ and $\alpha=1$, and $\gamma=1$ and $\alpha=10$. These sets represent a liquid Newtonian phase with a viscosity that is significantly smaller, equal, and significantly higher than the total viscosity of the suspending viscoelastic liquid, respectively.

We start by considering the limiting case of a spherical droplet, formally achieved by setting $Ca=0$ in Eqs.\eqref{eq:speedfinal} and \eqref{eq:shapefinal}. In this case, the up-down symmetry of the problem ensures that the expansion in Eq.\eqref{eq:speedfinal} only contains even powers of $Wi^{(A)}$; see also Eq.\eqref{eq:spherical4}.
Indeed, the speed of the droplet is expected to be independent of the direction of the external temperature gradient, thus requiring that all contributions to $U/U^{(0,0)}$ proportional to odd powers of $Wi^{(A)}$ vanish. (Note that the global sign of $U$ is set by the Newtonian velocity $U^{(0,0)}$.)  Eq.\eqref{eq:speedfinal}, which only captures the first two terms in this series, predicts that the speed of the droplet decreases with $Wi^{(A)}$ for any value of $\alpha$ and $\beta$. Since for $Ca=0$ the temperature field is given by its Newtonian profile, Eq.\eqref{eq:TNewt}, and since $\zeta(\theta) = 0$, we attribute the decrease in the droplet's speed to the effect of viscoelasticity in the outer fluid: the stresses generated by the polymer molecules slightly stretched from their equilibrium conformations by the fluid flow oppose the droplet's translational motion. The polymer-induced slow down increases with decreasing $\beta$, as the coupling between the fluid velocity and the polymeric stresses increases. While Eq.\eqref{eq:speedfinal} predicts that $U/U^{(0,0)}$ becomes negative at sufficiently large values of $Wi^{(A)}$, \emph{i.e.} that the droplet changes the direction of its motion, this conclusion is most likely an artefact of the low order of expansion employed in this work and would have to be corroborated by either the higher-order terms or numerical simulations.

\begin{figure}
\centering
\includegraphics[width=0.75\textwidth]{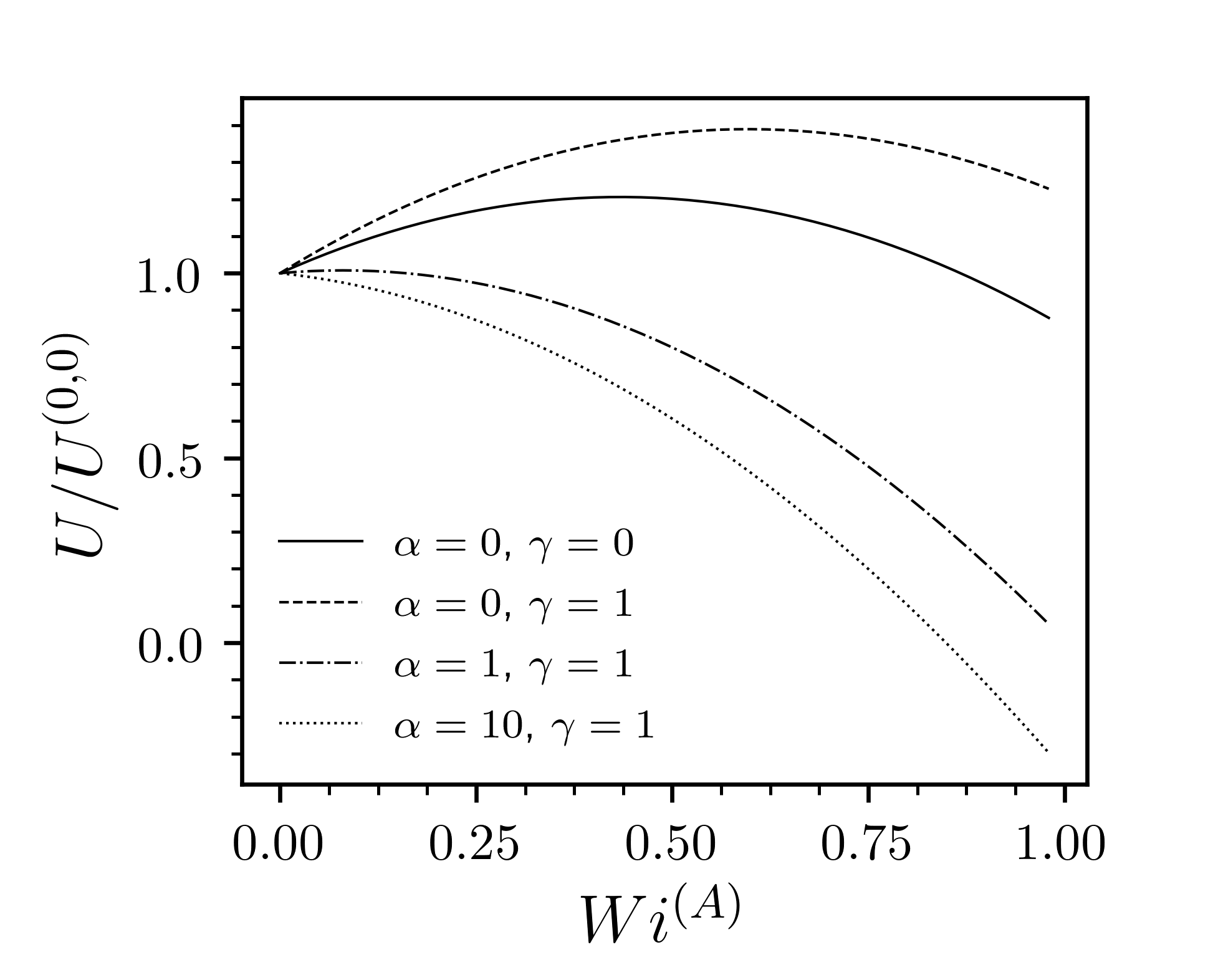}
\caption{The normalised droplet speed $U/U^{(0,0)}$ as a function $Wi^{(A)}$ for $Ca^{(A)} = 0.2$ and $\beta=0.1$. Although we expect our theory to be quantitatively correct only for $Wi^{(A)}<0.4$, here we also explore higher values of $Wi^{(A)}$ to stress the non-monotonic behaviour for small values of $\alpha$.}
\label{fig:UfixedCa}
\end{figure}

This behaviour changes for non-zero values of $Ca$. In Fig.\ref{fig:UfixedCa} we plot the prediction of Eq.\eqref{eq:speedfinal} for the four sets of parameters discussed above with $Ca^{(A)} = 0.2$ and $\beta=0.1$. In contrast to the spherical case, the droplet's speed is now a non-monotonic function of $Wi^{(A)}$ for most of the parameter values considered. For small values of $Wi^{(A)}$, the droplet speeds up in comparison to its Newtonian counterpart, while at larger values of $Wi^{(A)}$, it slows down again. When present, the speed up is caused by the $O(Ca\,Wi)$ term in Eq.\eqref{eq:speedfinal}, and we now analyse its physical origin.

To this effect, we consider an artificial problem of a deformed Newtonian droplet moving through a Newtonian outer fluid. We \emph{prescribe} a surface deformation in the form of $\zeta(\theta)=\Delta\,P_2(\cos\theta)$, where $\Delta$ is a small amplitude. We repeat the analysis discussed in Section \ref{sect:newtonian_solution} to first order in $\Delta$ and with $Wi=0$. Disregarding the normal-stress boundary condition, Eq.\eqref{eq:stressnormal}, which now defines an external force needed to create the deformation prescribed, we obtain for the droplet's speed
\begin{align}
\frac{U}{U^{(0,0)}} = 1 + \frac{2}{5} \Delta \left( \frac{33-18\alpha}{3(2+3\alpha)}+\frac{3(\gamma-1)}{2+\gamma}  \right).
\label{eq:Umisc}
\end{align}
We now identify $\Delta$ with the amplitude of the $O(Ca\,Wi)$ term in Eq.\eqref{eq:shapefinal}, and recover the $O(Ca\,Wi)$ term in Eq.\eqref{eq:speedfinal}. We, therefore, conclude, that once the $O(Wi)$ viscoelastic stresses deform the droplet, its speed up is a purely Newtonian effect. The structure of Eq.\eqref{eq:Umisc} suggests that it can be attributed to two mechanisms. When $\gamma=1$, the outer and inner fluids share the same thermal properties, and no changes to the temperature profile around a spherical droplet, Eq.\eqref{eq:TNewt}, will occur in response to the droplet deformation. Therefore, the first term proportional to $\Delta$ in Eq.\eqref{eq:Umisc} is associated with changes to the hydrodynamic resistance to the droplet's motion generated by the viscous stresses in both fluids, while the second term $O(\Delta)$ captures the response to changes in the fore-aft temperature gradient experienced by the droplet. This is further corroborated by the structure of the temperature profile inside the droplet,
\begin{align}
\left( \frac{3}{2+\gamma} + \frac{18(\gamma-1)}{5(2+\gamma)^2} \Delta \right) r \cos{\theta},
\end{align}
that changes compared to the Newtonian temperature profile, Eq.\eqref{eq:TNewt}, depending on whether $\gamma$ is larger or smaller than unity. We note that both effects can either speed the droplet up or slow it down compared to the Newtonian case, $Wi=0$, and that the speed up is observed for $\alpha < (16+17\gamma)/(21-3\gamma)$.

The $O(Ca\,Wi)$ term in Eq.\eqref{eq:shapefinal} is proportional to the second-order Legendre polynomial, $P_2(\cos\theta)$, and, therefore, describes a fore-aft symmetric deformation. Similar to the argument employed above, such deformations should be independent of the direction of motion and are thus described by even powers of the expansions parameters, such as $Ca\,Wi$. The higher order terms in Eq.\eqref{eq:shapefinal}, on the other hand, break the fore-aft symmetry, and are, therefore, coupled to odd total powers of the expansion parameters, such as $Ca^2\,Wi$ and $Ca\,Wi^2$. In general, the symmetry arguments require that the deformations described by even/odd Legendre polynomials be coupled to even/odd total powers of $Ca$ and $Wi$, respectively. 

\begin{figure}
\centering
\includegraphics[width=0.75\textwidth]{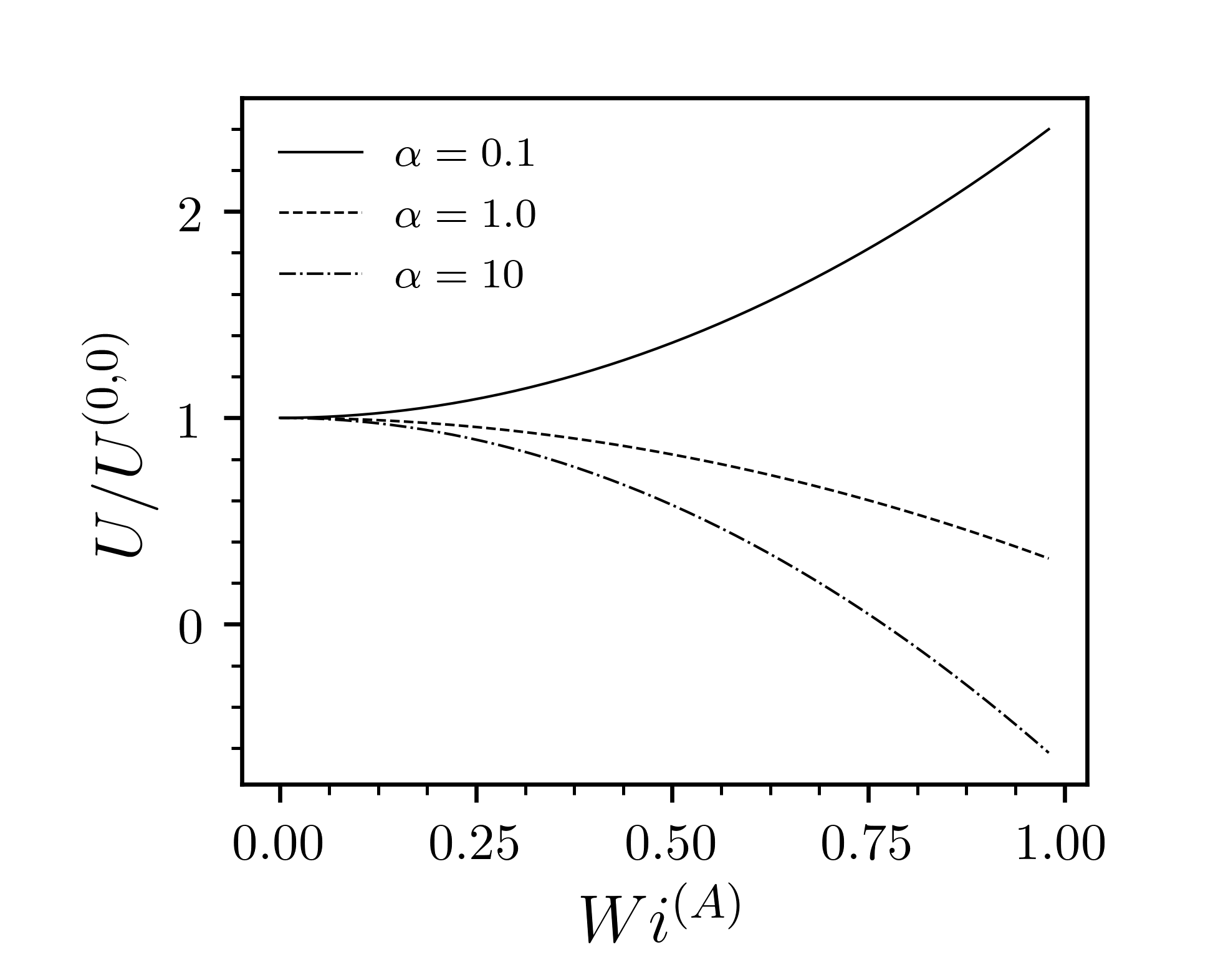}
\caption{The normalised droplet speed $U/U^{(0,0)}$ as a function $Wi^{(A)}$ for $\gamma=1$, $\beta=0.1$, and a fixed ratio $Ca/Wi=0.5$. }
\label{fig:UfixedCaDeRatio}
\end{figure}

Until now, we studied the predictions of Eqs.\eqref{eq:speedfinal} and \eqref{eq:shapefinal} as functions of $Wi^{(A)}$ for fixed values of $Ca^{(A)}$. However, in a typical experiment one is expected to vary the applied temperature gradient, thus simultaneously changing both the Weissenberg and Capillary numbers since both are proportional to $\nabla_\infty T$. To mimic such an experiment, in Fig.\ref{fig:UfixedCaDeRatio} we plot the prediction of Eq.\eqref{eq:speedfinal} as a function of $Wi^{(A)}$ for a fixed ratio $Ca/Wi$; the latter quantity is independent of the applied temperature gradient and is a function of the fluids' properties only. Unlike the fixed $Ca$ case studied above, Fig.\ref{fig:UfixedCaDeRatio} shows that the droplet's speed is monotonic in $Wi^{(A)}$, either increasing or decreasing depending on the values of $\alpha$ and $\gamma$. Eq.\eqref{eq:speedfinal} readily yields that the droplet's speed increases with $Wi$ as long as
\begin{align}
\frac{Ca}{Wi} > \frac{\frac{45(1+\alpha)(26+193\alpha)}{143(2+3\alpha)} +18(1-\beta)}{\left(22+13\alpha\right)\left( \frac{33-18\alpha}{3(2+3\alpha)} +\frac{3(\gamma-1)}{2+\gamma}\right)}.
\end{align}
For $\gamma=1$ and $\beta=0.1$, this condition is satisfied for $\alpha=0.1$, but not for $\alpha=1$ and $\alpha=10$, consistent with the behaviour in Fig.\ref{fig:UfixedCaDeRatio}. 

\begin{figure}
\centering
\includegraphics[width=0.95\textwidth]{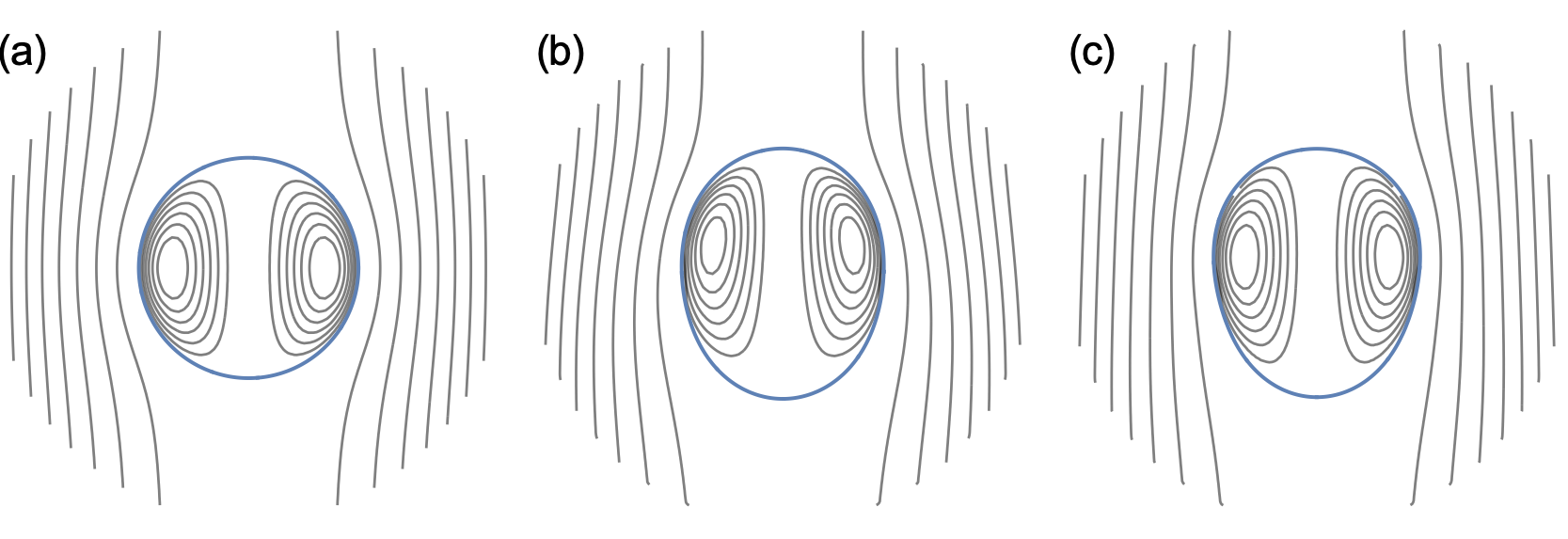}
\caption{Flow streamlines and the droplet shape for $\gamma=1$, $\beta=0.1$, and $Ca/Wi=0.5$. (a) Newtonian: $Wi^{(A)}=0.0$ and $\alpha=0.1$; (b) Viscoelastic: $Wi^{(A)}=0.3$ and $\alpha=0.1$; (c) Viscoelastic: $Wi^{(A)}=0.3$ and $\alpha=1.0$.}
\label{fig:streamlines}
\end{figure}

\begin{figure}
\centering
\includegraphics[trim=0 50 0 50, clip, width=0.95\textwidth]{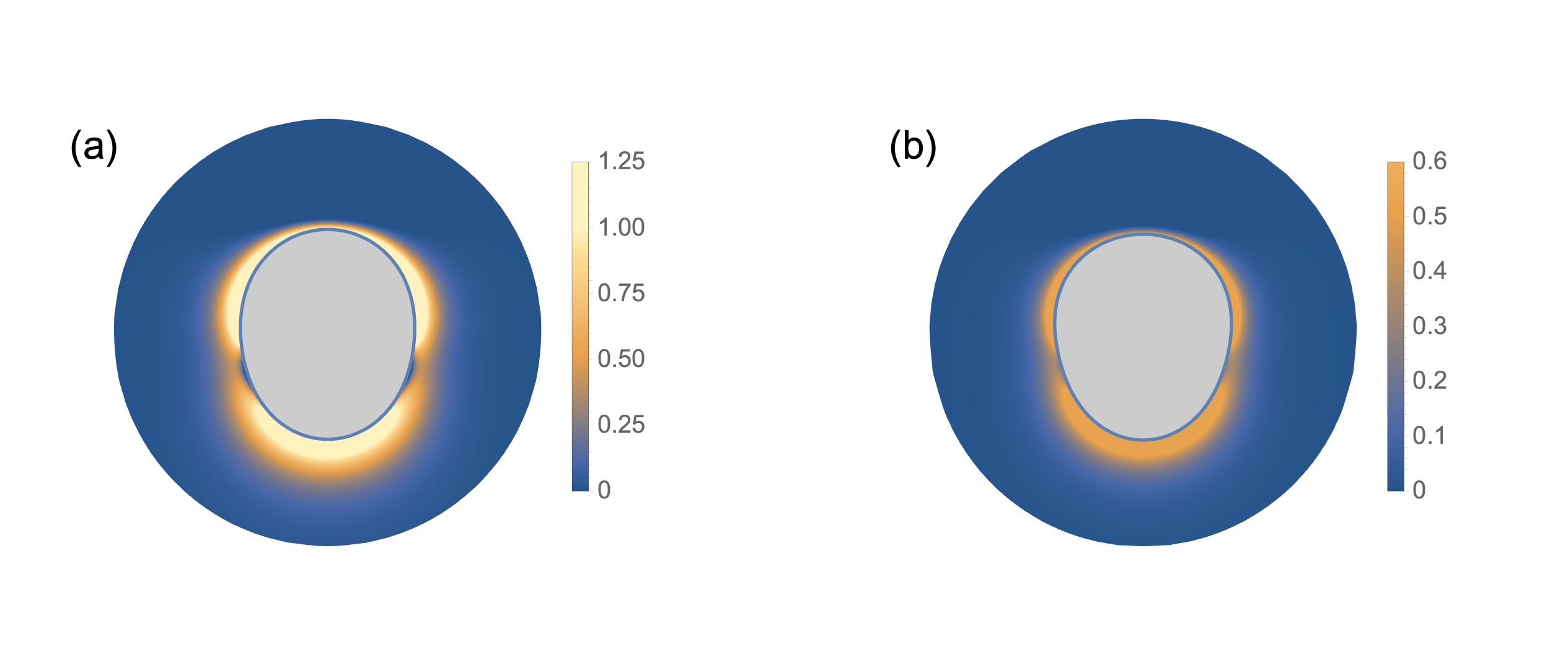}
\caption{Trace of the stress tensor in the outer fluid, $\mathrm{Tr}\,\bm{\tau}$,  for $\gamma=1$, $\beta=0.1$, $Ca/Wi=0.5$, and $Wi^{(A)}=0.3$. (a) $\alpha=0.1$; (b) $\alpha=1.0$.}
\label{fig:stressprofile}
\end{figure}

To gain further insight into the mechanical origins of the speed up/slowing down, we now closely inspect the case with $\gamma=1$, $\beta=0.1$, $Ca/Wi=0.5$ and $Wi^{(A)}=0.3$, with $\alpha$ either $0.1$ or $1.0$; the former case corresponds to a droplet moving faster than its Newtonian counterpart at the same conditions, while the latter case represents a smaller droplet speed. In Fig.\ref{fig:streamlines} we plot the flow streamlines inside and outside the droplet for these two cases, contrasted with the case of a Newtonian outer fluid, $Wi^{(A)}=0$, in Fig.\ref{fig:streamlines}(a), which exhibits the classical Hadamard-Rybczy\'{n}sky type toroidal vortex \citep{HADAMARD1911,rybczynski1911fortschreitende}. When the outer fluid is viscoelastic, these vortices become strongly asymmetric, indicating the loss of time-reversibility due to the viscoelastic memory effects, 
shifting towards the front part of the droplet, see Figs.\ref{fig:streamlines}(b) and (c). In Fig.\ref{fig:stressprofile} we show the trace of the stress tensor in the outer fluid, $\mathrm{Tr}\,\bm{\tau}$, for the same parameters; note that since the flow is incompressible, $\mathrm{Tr}\,\bm{\tau}$ only contains contributions from the polymeric part of the stress tensor, and is proportional to the local extension of polymer molecules \citep{Morozov2015}. In both cases, we observe a similar structure of the stress field, with strong polymer extension around the front and back stagnation points. As can be seen from  Fig.\ref{fig:stressprofile}, the thermocapillary interfacial stresses generate higher polymeric stresses in the case of a less viscous inner fluid, $\alpha=0.1$, as compared to the more viscous case, $\alpha=1$. When streaming along the curved leading surface of the droplet, the hoop stresses generated by the stretched polymers compress the droplet in the direction perpendicular to the direction of its motion thus extending it along the temperature gradient. This extension is larger in the less viscous case, leading to a larger droplet speed in this case. While the droplet stays relatively fore-aft symmetric in the less viscous case, it develops a significant fore-aft asymmetry for a more viscous inner fluid.

We conclude by comparing the predictions of Eqs.\eqref{eq:speedfinal} and \eqref{eq:shapefinal} against numerical simulations performed by \cite{CAPOBIANCHI20198}, who considered the current problem for vanishingly small Marangoni and Reynolds numbers, and matching fluid properties, $\gamma=1$ and $\alpha=1$. The Capillary number was fixed to $Ca=0.2$, leading to $Ca^{(A)}=0.027$.  In Table \ref{tab:comparison_vel} we present the dimensionless migration speed $U/U^{(0,0)}$ obtained from Eq.\eqref{eq:speedfinal} set against the simulations of \cite{CAPOBIANCHI20198} for $\beta=0.11$ and $\beta=0.5$. We observe that the numerical data differ from unity at $Wi=0$ due to inherent computational errors. This difference is about $1\%$ and sets the accuracy of the numerical data. For $Wi>0$, the difference between the two methods is consistently within 1\% -- 4\%, demonstrating a good predictive power of Eq.\eqref{eq:speedfinal} within its applicability range. When the Weissenberg number reaches $Wi=3.75$, the differences rise sharply, reaching $4\%$ for $\beta=0.5$, and $20\%$ for $\beta=0.11$. This is unsurprising since this value lies outside our estimate for the applicability range of Eq.\eqref{eq:speedfinal}, $Wi^{(A)}<0.4$.

\begin{table}
  \begin{center}
\def~{\hphantom{0}}
  \begin{tabular}{ccccccc}
&&  \multicolumn{2}{c}{$\beta=0.5$} && \multicolumn{2}{c}{$\beta=0.11$}\\
\cline{3-4}\cline{6-7}\\
$Wi$ &$Wi^{(A)}$ & $(U/U^{(0,0)})_{an}$ & $(U/U^{(0,0)})_{num}$ && $(U/U^{(0,0)})_{an}$& $(U/U^{(0,0)})_{num}$\\
\hline
0& 		0&		1.0000&	0.9931&&		1&		0.9930 \\
0.1875& 	0.025&	1.0000&	0.9930&&		0.9999&	0.9952 \\
0.375& 	0.05&	0.9993&	0.9939&&		0.9983&	0.9900 \\
0.75& 	0.1&		0.9959&	0.9897&&		0.9909&	0.9881\\
1.5& 		0.2&		0.9809&	0.9604&&		0.9584&	0.9540 \\
2.25& 	0.3&		0.9548&	0.9440&&		0.9027&	0.9280\\
3.75& 	0.5&		0.8699&	0.9010&&		0.7215&	0.8650
  \end{tabular}
  \caption{Comparison between the dimensionless migration speed observed in numerical simulations of \cite{CAPOBIANCHI20198}, $(U/U^{(0,0)})_{num}$, and the predictions of Eq.\eqref{eq:speedfinal}, $(U/U^{(0,0)})_{an}$.}
  \label{tab:comparison_vel}
  \end{center}
\end{table}

\begin{figure}
\centering
\includegraphics[width=0.95\textwidth]{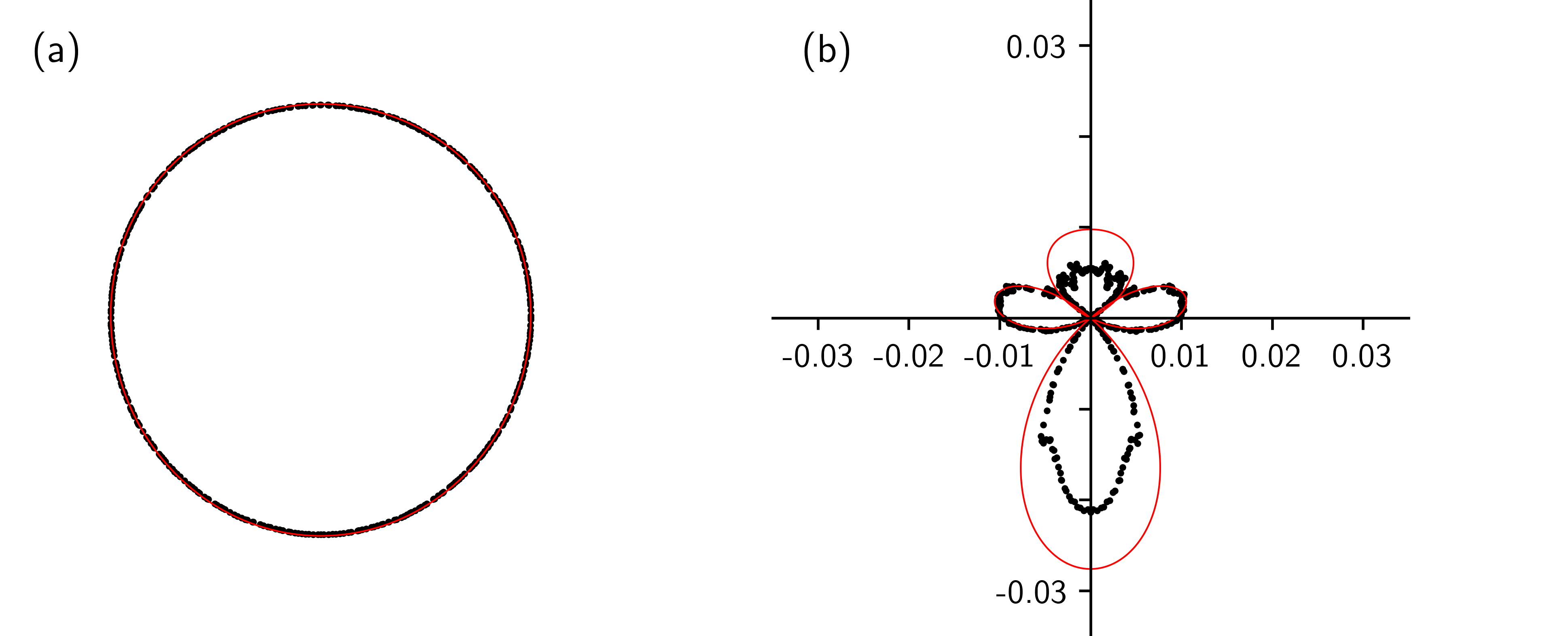}
\caption{Comparison between the droplet shape obtained by \cite{CAPOBIANCHI20198} (black circles) and the prediction of Eq.\eqref{eq:shapefinal} (red line) for $\gamma=1$, $\alpha=1$, $Ca=0.2$, $Wi=2.25$, and $\beta=0.11$. (a) The full shape, $1+\zeta(\theta)$; (b) the corresponding deviation from the spherical shape, $\zeta(\theta)$.}
\label{fig:shapes}
\end{figure}

In addition, we  compare the steady-state droplet shape determined numerically with the ones predicted by the present analytical calculations Eq.\eqref{eq:shapefinal}. In Fig.\ref{fig:shapes}(a), we show the polar plot of the droplet shape, $1+\zeta(\theta)$, compared to the results of  \cite{CAPOBIANCHI20198}, for a representative value of $Wi=2.25$ with $\beta=0.11$. The two shapes differ only slightly from each other and from a spherical droplet. To stress the differences, in Fig.\ref{fig:shapes}(b) we plot the deviation from the spherical shape, $\zeta(\theta)$, for the same parameters. While there is a good overall agreement in the symmetry of $\zeta(\theta)$, quantitative comparison is lacking around the front and back stagnation points. As discussed above, the polymer stresses accumulate strongly around these areas, see Fig.\ref{fig:stressprofile}, and the local Weissenberg numbers can become significantly larger than the globally prescribed value $Wi^{(A)}$, thus exceeding the applicability range of Eq.\eqref{eq:shapefinal}. This is further corroborated by the droplet shape observed by  \cite{CAPOBIANCHI20198} for $Wi=3.75$, which developed a cusp at the rear stagnation point due to a very strong stress localisation. Finally, we note that we could not perform a meaningful shape comparison between our theory and simulations of \cite{CAPOBIANCHI20198}  at lower values of $Wi$, where a better agreement is expected. At those conditions, the deviation from the spherical shape, $\zeta(\theta)$, is much smaller than for the case shown in Fig.\ref{fig:shapes}(b), and the accuracy of the numerical data is insufficient to resolve it. However, the droplet speed, which can be seen as a proxy for the shape, is in a good agreement with the numerical data, as discussed above, and we conclude that the two approaches are in a semi-quantitative agreement within the applicability range of the current theory, $Wi^{(A)}<0.4$.

\section{Conclusions}
\label{sect:conclusions}

In this work we considered theoretically the problem of a Newtonian droplet moving in an otherwise quiescent infinite viscoelastic fluid under the influence of an externally applied temperature gradient. The outer fluid was described by the Oldroyd-B model, and the problem was solved for small Weissenberg and Capillary numbers in terms of a double perturbation expansion. The analysis was conducted assuming the absence of gravity and negligible convective transport effects.
The main results of our work, Eqs.\eqref{eq:speedfinal} and \eqref{eq:shapefinal}, give predictions for the droplet speed and  shape as a function of the fluids' parameters. In the absence of the shape deformation, $Ca=0$, the droplet speed decreased monotonically for sufficiently viscous inner fluids, while for fluids with a smaller viscosity ratio $\alpha$, the droplet speed first increased and then decreased as a function of the Weissenberg number. For small but finite values of the Capillary number, the droplet speed behaved monotonically as a function of the applied temperature gradient for a fixed $Ca/Wi$ ratio. We demonstrated that this behaviour is related to the polymeric stresses deforming the droplet in the direction of its migration, while the associated changes in its speed were Newtonian in nature, being related to a changes in the droplet's hydrodynamic resistance and its internal temperature distribution. When compared to the results of numerical simulations, our theory exhibited a good predictive power within its applicability range, i.e. for sufficiently small values of $Wi$ and $Ca$.
The problem of thermocapillary motion of droplets in viscoelastic fluids and the results presented here can be of potential interest to the space manufacturing sector, and in microfluidic applications, where the small characteristic lengths scale would allow thermocapillary effects to prevail with respect to buoyancy.
 
 \section*{Acknowledgements}
RJP gratefully acknowledges funding from the EPSRC (UK) through grant no. EP/M025187/1.

For the purpose of open access, the authors have applied a Creative Commons Attribution (CC BY) licence to any Author Accepted Manuscript version arising from this submission.

\section*{Declaration of interests}
The authors report no conflict of interest.

 \bibliographystyle{elsarticle-num} 
 \bibliography{cas-refs}





% \end{thebibliography}
\end{document}